\documentclass[twocolumn]{emulateapj}  
\usepackage{epstopdf}
\usepackage{ulem}
\usepackage{amssymb}
\usepackage{natbib}
\usepackage{times}
\usepackage{graphicx,bm,amssymb}
\usepackage{pstricks}

\usepackage{psfrag}
\usepackage[colorlinks=true,linkcolor=blue,citecolor=blue]{hyperref}
\newcommand{\be}{\begin{equation}}
\newcommand{\ee}{\end{equation}}
\newcommand{\bse}{\begin{subequations}}
\newcommand{\ese}{\end{subequations}}
\newcommand{\bary}{\begin{eqnarray}}
\newcommand{\eary}{\end{eqnarray}}

\shorttitle{}
\shortauthors{Fraija N.}

\begin{document}
\title{Modeling the early multiwavelength emission in GRB130427A}
\author{N. Fraija\altaffilmark{1}, W. Lee\altaffilmark{1} and P. Veres\altaffilmark{2} }
\affil{$^1$Instituto de Astronom\'ia, Universidad Nacional Aut\'{o}noma de M\'{e}xico, Apdo. Postal 70-264, Cd. Universitaria, DF 04510, M\'{e}xico}
\affil{$^2$Center for Space Plasma and Aeronomic Research (CSPAR), University of Alabama in Huntsville, Huntsville, AL 35899, USA}
\email{nifraija@astro.unam.mx, wlee@astro.unam.mx, pv0004@uah.edu}
\date{\today}
\begin{abstract}
One of the most powerful gamma-ray bursts, GRB 130427A was swiftly detected from GeV $\gamma$-rays to optical wavelengths.  In the GeV band, the Large Area Telescope (LAT) on board the Fermi Gamma-Ray Space Telescope observed the highest-energy photon ever recorded of 95 GeV, and a bright peak in the early phase followed by  emission temporally extended  for more than 20 hours.  In the optical band, a bright flash with a magnitude of $7.03\pm 0.03$ in the time interval from 9.31~s to 19.31~s after the trigger was reported by RAPTOR in r-band.   We study the origin of the GeV $\gamma$-ray emission, using the multiwavelength observation detected in X-ray and optical bands. The origin of the temporally extended LAT, X-ray and optical flux is naturally interpreted as synchrotron radiation and the 95-GeV photon and  the integral flux upper limits placed by the HAWC observatory are consistent  with  synchrotron self-Compton from an adiabatic forward shock propagating into the stellar wind of its progenitor.  The extreme LAT peak and the bright optical flash are explained through synchrotron self-Compton and synchrotron emission from the reverse shock, respectively, when the ejecta  evolves in thick-shell regime and carries a significant magnetic field. 
\end{abstract}
\section{Introduction}
Gamma-ray bursts (GRBs) are the most luminous explosions in the universe. Based on photometric and spectroscopic observations, long GRBs (lGRBs) have usually been associated  to the core collapse of massive stars leading to supernovae (CCSNe) of type Ib, Ic or II  \citep{2006ARA&A..44..507W, 2012grbu.book..169H, 2003Natur.423..847H}.  In the cosmological scenario, the large isotropic energy release (up to  $\sim 10^{55}$ erg),   short variability timescale (down to $\sim10^{-3}$ s), and nonthermal gamma-ray spectra leads to an ultrarelativistic expansion with a large bulk Lorentz factor in the range of $10^2 - 10^3$. In the standard fireball model,  the expanding relativistic ejecta interacts with the surrounding medium generating reverse and forward shocks.  The long-lasting forward shock (FS) leads to a continuous softening of the afterglow spectrum \citep{2007MNRAS.379..331P,2004MNRAS.353..647N}, whereas the reverse shock (RS) that propagates into the ejecta gives rise to a strong and short peak.   After the peak, no new electrons are injected and the material cools adiabatically, although if the central engine emits slowly-moving material the RS could survive from hours to days \citep{2007MNRAS.381..732G, 2007ApJ...665L..93U}.\\ 
%
%
Due to its intensity and proximity (z=0.34; \citet{2013GCN..14686...1L}), one of the most energetic bursts,  GRB 130427A,  was observed in GeV-MeV $\gamma$-rays, X-rays and the optical band.  GRB 130427A was detected on 2013 April 27 at 07:47:06.42 UTC by the Gamma-ray Burst Monitor (GBM)  on board Fermi   \citep{2013GCN..14473...1V} and afterwards by several orbiting satellites and multiple ground-based telescopes \citep{2014Sci...343...48M, 2014Sci...343...42A, 2013GCN..14484...1P}. The Large Area Telescope (LAT) observed this burst for $\sim$ 70~ks exhibiting a bright peak at $\sim$ 15~s after the GBM trigger. The optical and LAT emission showed a  close correlation during the first 7000~s \citep{2014Sci...343...38V}.  In particular,  a bright optical flash peaking at 15~s  was temporally correlated with the LAT peak.\\
%
Some authors have claimed that the multiwavelength afterglow observed in GRB 130427A,  from dozens of seconds to days after the GBM trigger,  can be modeled as synchrotron emission of relativistic electrons accelerated in the stellar wind of the standard reverse and forward shock \citep{2013ApJ...776..119L, 2014ApJ...781...37P, 2013ApJ...779L...1K, 2014Sci...343...48M}. For instance, the bright optical flash was better fitted by reverse shock emission \citep{2014Sci...343...38V}.  Other sets of models have interpreted the temporally extended Fermi-LAT flux  through the synchrotron radiation, Compton scattering emission and electromagnetic cascades induced by ultrarelativistic hadrons \citep{2014Sci...343...42A, 2013ApJ...776...95F, 2013ApJ...773L..20L}.\\
%
%
%
Recently, we have presented a leptonic model based on an early stellar-wind afterglow  to describe the temporally extended LAT, X-ray and optical fluxes, as well as  the brightest peak  present in the LAT light curve of GRB 110731A \citep{2015ApJ...804..105F}.  In this paper, we apply this model to explain the multiwavelength afterglow observations of GRB 130427A: the bright LAT peak  and optical flash by reverse shock emission and the temporally extended LAT, X-ray and optical fluxes by forward shock emission.  The paper is arranged as follows: in Section 2 we give a brief description of GRB 130427A observations; in Section 3 we present a leptonic model  based on external shocks (forward and reverse) that evolve adiabatically in a stellar wind with the quantities observed in GRB 130427A;  in section 4  we discuss our results, and brief conclusions are given in section 5.
\section{GRB 130427A}
GRB 130427A triggered the Gamma-ray Burst Monitor (GBM)  on board the Fermi satellite at 07:47:06.42 UTC on 2013 April 27 \citep{2013GCN..14473...1V}. Promptly, the Burst Alert Telescope (BAT) on board Swift triggered on the ongoing burst at 07:47:57.51 UTC.  The structure of the light curve (LC) revealed by the BAT instrument in the 15- to 350-keV band showed a complex structure with a duration of $\sim$ 20~s.   The Swift Ultra Violet Optical Telescope (UVOT) began observations at $\sim$ 181~s, whereas observations with the Swift X-ray Telescope (XRT) started at  $\sim$ 195~s  \citep{2014Sci...343...48M}.  Its exceedingly bright prompt emission was also detected by other satellites (SPI-ACS/INTEGRAL; \citep{2013GCN..14484...1P}   AGILE; \cite{2013GCN..14515...1V}, Konus-Wind; \cite{2013GCN..14487...1G}, NuSTAR; \cite{2013ApJ...779L...1K}   RHESSI; \cite{2013GCN..14590...1S}) and multiple ground- and space follow-up facilities (MAXI/GSC; \cite{2013GCN..14462...1K}, VLT/X-shooter; \cite{2013GCN..14491...1F}).  For instance, optical  spectroscopy from Gemini-North found the redshift of the GRB to be z=0.34 (confirmed later by VLT/X-shooter; \cite{2013GCN..14491...1F}), revealing the closeness to Earth \citep{2013GCN..14686...1L} and the  optical/near infrared (NIR) counterpart observed with the Hubble Space Telescope suggested the association of GRB 130427A with a Type Ic supernova (SN2013cq) \citep{2013GCN..14686...1L, 2013ApJ...776...98X}.  RAPTOR (Rapid Telescope for Optical Response) reported on the bright optical flash with a magnitude of $7.03\pm 0.03$ in the time interval from 9.31 s to 19.31 s after the GBM trigger \citep{2014Sci...343...38V}. After the peak, the flash faded with a power law flux decay with index $\alpha=-1.67\pm 0.07$ and was detected for $\sim$ 80 s until it faded below the $\sim 10$th magnitude sensitivity limit of the RAPTOR full-sky monitors. LAT followed-up this burst until it became occulted by the Earth 715 s  after the GBM trigger. The burst emerged at 3.1 ks  and was detected for $\sim$ 20 hr, only interrupted by further occultations \citep{2014Sci...343...42A}.  This burst presented the highest fluence with isotropic energy $\sim 1.4\times 10^{54}$ erg and the highest energy photons ever detected, 73 GeV and 95 GeV  observed at 19 s and 244 s, respectively.  For a single power-law fit to the energy flux light curve, the Fermi Collaboration reported a temporal index of $-1.17\pm 0.06$, consistent with other Fermi-LAT bursts \citep{2013ApJS..209...11A}.  Finally,  TeV $\gamma$-ray observatories such as the High Altitude Water Cherenkov observatory (HAWC; \cite{2013GCN..14549...1L, 2015ApJ...800...78A}) and the Very Energetic Radiation Imaging Telescope Array System (VERITAS; \cite{2014ApJ...795L...3A})   followed up observations. Although no statistically significant excess of counts was  registered  by these TeV observatories, upper limits were placed on the emission.\\
Given some similarities,  such as the presence of a temporally LAT extended emission longer than the duration of the prompt emission and a bright LAT peak in coincidence with the prompt phase between the  bursts GRB130427 and GRB110931A, we summarize in  Table~1  the relevant observational quantities.  
\begin{center}
\begin{center}
\scriptsize{\textbf{Table 1. Observed quantities for  GRB110731A and GRB130427A.}}\\
\end{center}
\begin{tabular}{ l c c }
  \hline \hline
 \scriptsize{Parameter} & \scriptsize{GRB110731A} & \scriptsize{GRB130427A} \\
 \hline
\scriptsize{Isotropic Energy ($\times 10^{54}$ erg)}    & \scriptsize{ 0.76 }  &  \scriptsize {0.96} \\
\scriptsize{Redshift}  & \scriptsize{2.83}  &  \scriptsize{$0.34$ }\\
\scriptsize{Duration of prompt emission (s)}    & \scriptsize{$\sim\,14^a$ }  &  \scriptsize{ $\sim 138^b$}\\
\scriptsize{Peak time (s)} & \scriptsize{$\sim$ 5.5} & \scriptsize{$\sim$ 15 }\\
\scriptsize{Duration of bright peak (s)}    & \scriptsize{ $\sim\,1$ }  &  \scriptsize{$\sim 9$}\\
\scriptsize{Duration of extended emission (s)}    & \scriptsize{$\sim\,10^3$ }  &  \scriptsize{ $> 10^4$}\\
\scriptsize{Prompt GeV emission (erg/cm$^2$)}    & \scriptsize{0.47 $\times10^{-4}$ }  &  \scriptsize{ $\sim\, 10^{-4}$}\\
\scriptsize{Highest energy photon (GeV)}    & \scriptsize{$3.4\, (at \sim\,436$ s)}  &  \scriptsize{$95\, (at \sim\,244$ s)}\\
\scriptsize{Main references}    & \scriptsize{$1$}  &  \scriptsize{$2, 3, 4$}\\
\hline
\end{tabular}
\end{center}
\begin{flushleft}
\scriptsize{
Notes.\\
$^a$ Most of energy was released in the first $\sim$ 7 s.\\
$^b$  Most of energy was released in the first $\sim$ 18 s.\\
\textbf{References}. (1) \citet{2013ApJ...763...71A}; (2) \cite{2014Sci...343...42A}; (3) \cite{2014Sci...343...38V}; (4) \cite{2013GCN..14686...1L}.} 
\end{flushleft}
\vspace{1cm}
\section{External Shock Model}
As the ultrarelativistic blast wave spreads into the stellar dense wind of the progenitor, it is decelerated leading to forward and reverse shocks. The  afterglow dynamics will depend on its mass and in some cases, the emission processes (synchrotron and/or Compton scattering) generated at internal and external shocks which could be simultaneously present in the light curve \citep{2007MNRAS.379..331P,2004MNRAS.353..647N, 2003ApJ...597..455K, 2003MNRAS.342.1131W, 1997ApJ...476..232M}.  We hereafter use primes (unprimes) to define the quantities in a comoving (observer) frame, the universal constants   c=$\hbar$=1 in natural units and the values of cosmological parameters $H_0=$ 71 km s$^{-1}$ Mpc$^{-1}$, $\Omega_m=0.27$, $\Omega_\lambda=0.73$  \citep{2003ApJS..148..175S}. The subscripts f and r refer throughout this paper to the FS and RS, respectively and the convention $Q_x=Q/10^x$ will be adopted in c.g.s. units.
\subsection{Forward Shocks}
Afterglow hydrodynamics involves a relativistic blast wave expanding into the medium with density
\be\label{rho}
\rho=A_f\,r^{-2} \hspace{0.3cm} {\rm with}  \hspace{0.3cm}  A_f=\frac{\dot{M}_w}{4\pi V_w}\,,
\ee
where $\dot{M}_w$ is the mass loss rate and $V_w$ is the wind velocity.  Requiring the observable quantities: isotropic energy $E=1.4\times10^{54}$ erg \citep{2014Sci...343...42A}, redshift  $z=0.34$ \citep{2013GCN..14686...1L,2013GCN..14491...1F}, the stellar wind density  $A_f=A_{\star,f}\,(5.0\times 10^{11})$ g/cm \citep{2000ApJ...536..195C, 2013ApJ...776..119L, 2014ApJ...781...37P, 2013ApJ...779L...1K},  the bulk Lorentz factor $\Gamma_f= \Gamma_{\star,f}\,10^2$ and the index of power-law distribution of accelerated electrons $p=2.2$, we will apply the leptonic model developed in \citet{2015ApJ...804..105F}.  The value of  this power index was obtained linking the relation of synchrotron flux ($F_\nu\propto t^{-\alpha}\nu^{-\beta}$) with  the observed slopes of temporal decays of GeV $\gamma$-ray ($\alpha_{GeV}=-1.17\pm0.06$; \cite{2014Sci...343...42A}), X-ray ($\alpha_{X}=-1.29^{+0.02}_{-0.01}$; \cite{2014Sci...343...48M}) and optical ($\alpha_{opt}=-1.67\pm 0.07$; \cite{2014Sci...343...38V}) fluxes.  For an ultra relativistic and adiabatic blast wave, the deceleration time  is
\be\label{t_dec}
t_{dec}\simeq674.2\,{\rm s}\biggl(\frac{1+z}{1.34}\biggr)\,\xi^{-2}_{-0.3}\,E_{54.3}\,A^{-1}_{\star,f}\,\Gamma^{-4}_{\star,f}\,,
\ee
where the estimated values of $\xi$ for this case are $\sim$ 1 (low energy) and $\sim$ 0.5 (high energy)  \citep{1998ApJ...493L..31P, 2000ApJ...536..195C}.
\paragraph{Synchrotron emission}. 
Considering that electrons are accelerated to a power-law distribution $N(\gamma_e) d\gamma_e\propto \gamma_e^{-p} d\gamma_e$  and the energy density is equipartitioned to accelerate electrons and to amplify/create the magnetic field through the micro physical parameters $\epsilon_{e,f}$ and $\epsilon_{B,f}$, respectively, the  e-minimum Lorentz factor and the magnetic field can be written as
\be\label{gamma_m}
\gamma_{e,m,f}= 3.1\times 10^4\, \epsilon_{e,f}\,\Gamma_{\star,f}\,,
\ee
and
{\small
\be\label{Bf}
B'_f\simeq 6.6 \times 10^3\, {\rm G} \biggl(\frac{1+z}{1.34}\biggr)^{1/2}\xi^{-1}_{-0.3}\, \epsilon_{B,f}^{1/2}\,\Gamma_{\star,f}\,E^{-1/2}_{54.3}\,t^{-1/2}_{1}\,A_{\star,f}\,,
\ee
}
respectively.    When  the expanding relativistic ejecta encounters the stellar wind, it starts to be decelerated, then electrons are firstly heated and after cooled down  by synchrotron emission. Comparing the deceleration time scale (eq. \ref{t_dec}) with the cooling  {\small $t_{e,syn}\simeq 3m_e/(16\sigma_T)\,(1+x_f)^{-1}\,(1+z)\,\epsilon^{-1}_{B,f}\,\rho^{-1}\,\Gamma^{-3}_f\,\gamma_e^{-1}$} and the acceleration $t_{acc}\simeq \frac{2\pi\,m_e}{q_e}(1+z)\,\Gamma^{-1}_f\,{B'}^{-1}_f \gamma_e$ time scales for synchrotron radiation,  then the cooling and maximum Lorentz factors  are {\small$\gamma_{e,c,f}=\frac{3m_e\xi^4}{\sigma_T}\,(1+x_f)^{-1}\,(1+z)^{-1}\,\epsilon^{-1}_{B,f}\,\Gamma_f\,A^{-1}_f\,t$} and {\small $\gamma_{e,max,f}\simeq \sqrt{\frac{9\sqrt2\,q_e}{16\,\pi \sigma_T}}\,\xi^{1/2}\,(1+z)^{-1/4}\epsilon_{B,f}^{-1/4}\Gamma^{-1/2}_f\,E^{1/4}\,A^{-1/2}_{f}\,t^{1/4}$}, respectively. Here $\sigma_T$ is the Thomson cross section, $q_e$ is the elementary charge and the term $(1+x_f)$ is introduced because a once-scattered synchrotron photon generally has energy larger than the electron mass in the rest frame of the second-scattering electrons \citep{2001ApJ...548..787S}.\\
Considering the electron Lorentz factors ($\gamma^2_{e,i,f}$  for i=m,c and max) and  eqs. (\ref{t_dec}) and (\ref{gamma_m}), the synchrotron spectral breaks  computed through the synchrotron emission $E_{i,f}=\frac{q_e}{m_e}\,(1+z)^{-1}\,\Gamma_f\,B'\gamma^2_{e,i,f}$  can be written as
\vspace{0.8cm}
{\small
\bary\label{synforw_a}
E^{syn}_{\rm \gamma,a,f} &\simeq&  2.1\times 10^{-1}\,{\rm eV}\, \biggl(\frac{1+z}{1.34}\biggr)^{-2/5}\xi^{-6/5}_{-0.3}\,\epsilon_{e,f}^{-1}\,\epsilon_{B,f}^{1/5}\,A^{6/5}_{\star,f}\cr
&&\hspace{4.3cm}\times\,E^{-2/5}_{54.3}\,  t_{1}^{-3/5}\cr
E^{syn}_{\rm \gamma,m,f} &\simeq&  51.7\,{\rm MeV}\, \biggl(\frac{1+z}{1.34}\biggr)^{1/2}\xi_{-0.3}^{-3}\,\epsilon_{e,f}^2\,\epsilon_{B,f}^{1/2}\,E^{1/2}_{54.3}\,  t_{1}^{-3/2}\cr
E^{syn}_{\rm \gamma,c,f}  &\simeq&  7.3\times 10^{-6}\, {\rm eV}\, \biggl(\frac{1+z}{1.34}\biggr)^{-3/2}\xi_{-0.3}^{5}\,(1+x_f)^{-2}\, \epsilon_{B,f}^{-3/2}\cr
&&\hspace{4.2cm}\times\,A^{-2}_{\star,f}\, E^{1/2}_{54.3}\, t_{1}^{1/2}\, \cr
E^{syn}_{\rm \gamma,max,f}  &\simeq&61.9\, {\rm GeV} \biggl(\frac{1+z}{1.34}\biggr)^{-3/4}\xi_{-0.3}^{-1/2}E^{1/4}_{54.3}\,A^{-1/4}_{\star,f}t^{-1/4}_1\,.
\eary
}
\noindent The synchrotron self-absorption energy $E^{syn}_{\rm \gamma,a,f}$ was calculated through the absorption coefficient $\alpha_{\rm \epsilon'_a}$ \citep{1986rpa..book.....R} and the condition $\alpha_{\rm \epsilon'_a} r/\Gamma=1$ \citep{1999ApJ...523..177W,1999ApJ...527..236G}.  The maximum synchrotron flux  $F^{syn}_{\rm \gamma,max,f}=N_e P_{\nu,max}/4\pi D^2$ given as a function of the peak spectral power  {\small $P_{\nu,max}\simeq  \sigma_T(m_e/3q_e)\,(1+z)^{-1}\,\Gamma_f\,B'_f$}  can be explicitly written as 
\bary\label{Fsyn}
F^{syn}_{\rm \gamma,max,f}&\simeq& 7.1\times 10^{5}\,{\rm mJy} \biggl(\frac{1+z}{1.34}\biggr)^{3/2}\xi_{-0.3}^{-1}\,\epsilon_{B,f}^{1/2} \,A_{\star,f}\,D^{-2}_{28}\cr
&&\hspace{4.1cm}\times\,E^{1/2}_{54.3}\,t^{-1/2}_{1}\,,
\eary
where $D$ is the luminosity distance from the source. \\
Using the synchrotron spectral breaks (eq. \ref{synforw_a}) and synchrotron spectra \citep{1998ApJ...497L..17S},  the LC in the fast-cooling regime is
{\small
\begin{eqnarray}
\label{fcsyn_t}
[F_\nu]^{syn}= \cases{ 
F^{syn}_{\nu,fl}\,\,\,,\hspace{1cm} E^{syn}_{\rm \gamma,c,f}<E^{syn}_\gamma<E^{syn}_{\rm \gamma,m,f}, \cr
F^{syn}_{\nu,fh}\,\,\,,\hspace{1cm}E^{syn}_{\rm \gamma,m,f}<E^{syn}_\gamma<E^{syn}_{\rm \gamma,max,f}. \cr
}
\end{eqnarray}
}
where  $F^{syn}_{\nu,fh}$ is
{\small
\bary\label{A1}
F^{syn}_{\nu,fh}&=&  2.8\times 10^{-1}\,\,  {\rm mJy}\,(1+x_f)^{-1}\biggl(\frac{1+z}{1.34}\biggr)^{\frac{p+2}{4}}\xi_{-0.3}^{3(1-\frac{p}{2})} \cr
&&\times\,  \epsilon_{e,f}^{p-1}\,\epsilon_{B,f}^{\frac{p-2}{4}}\,E_{54.3}^{\frac{p+2}{4}}\,D_{28}^{-2} \,t_1^{-\frac{3p-2}{4}}\,\left(\frac{E^{syn}_{\rm \gamma,f}}{100\,{\rm MeV}}\right)^{-\frac{p}{2}}\,, 
\eary
}
\noindent and $F^{syn}_{\nu,fl}$ is given in \cite{2015ApJ...804..105F}. The LC in the slow-cooling regime is
{\small
\begin{eqnarray}
\label{scsyn_t}
[F_\nu]^{syn}=\cases{
F^{syn}_{\nu,sl}\,\,\,,\hspace{1cm} E^{syn}_{\rm \gamma,m,f}<E^{syn}_\gamma<E^{syn}_{\rm \gamma,c,f},\cr
F^{syn}_{\nu,sh}\,\,\,,\hspace{1cm} E^{syn}_{\rm \gamma,c,f}<E^{syn}_\gamma<E^{syn}_{\rm \gamma,max,f}\,, \cr
}
\end{eqnarray}
}
with $F^{syn}_{\nu,sh}$ and $F^{syn}_{\nu,sl}$ given by
{\small
\bary\label{A2}
F^{syn}_{\nu,sh}&=&  6.9\times 10^3\,\, {\rm mJy}\,(1+x_f)^{-1}\biggl(\frac{1+z}{1.34}\biggr)^{\frac{p+2}{4}}\xi_{-0.3}^{3(1-\frac{p}{2})} \cr
&&\times\,  \epsilon_{e,f}^{p-1}\,\epsilon_{B,f}^{\frac{p-2}{4}}\,E_{54.3}^{\frac{p+2}{4}}\,D_{28}^{-2} \,t_1^{-\frac{3p-2}{4}}\,\left(\frac{E^{syn}_{\rm \gamma,f}}{10\,{\rm keV}}\right)^{-\frac{p}{2}}\,,  
\eary
}
and
{\small
\bary\label{A3}
F^{syn}_{\nu,sl}&\simeq& 2.8\times 10^9\,\, {\rm mJy}\,\biggl(\frac{1+z}{1.34}\biggr)^{\frac{p+5}{4}}\xi_{-0.3}^{\frac{(1-3p)}{2}} \epsilon_{e,f}^{p-1}\,\epsilon_{B,f}^{\frac{p+1}{4}}\,A_{\star,f}\cr
&&\hspace{1.5cm}\times\, E_{54.3}^{\frac{p+1}{4}}\,D_{28}^{-2}\,t_1^{-\frac{3p-1}{4}}\, \left(\frac{E^{syn}_{\rm \gamma,f}}{2\,{\rm eV}}\right)^{\frac{1-p}{2}}\,,
\eary
}
respectively. The transition time ($t^{syn}_0$) from fast- to slow-cooling spectrum is 
{\small
\bary\label{t0}
t^{syn}_0=5.9\times10^7 {\rm s}\,\biggl(\frac{1+z}{1.34}\biggr)\xi_{-0.3}^{-4}\epsilon_{e,f}\,\epsilon_{B,f}\,A_{\star,f}\,.
\eary
}
\paragraph{SSC emission}
Fermi-accelerated electrons can scatter synchrotron photons up to higher energies $E^{ssc}_{\gamma,i}\simeq 2 \gamma^2_{e,i} E^{syn}_{\gamma,i}$. From the synchrotron spectral breaks (eq. \ref{synforw_a}),  the SSC spectral breaks are
{\small
\bary\label{sscforw_a}
E^{ssc}_{\rm \gamma, m,f} &\simeq&  1.2\times 10^6\,{\rm TeV}\biggl(\frac{1+z}{1.34}\biggr)\xi_{-0.3}^{-4}\,\epsilon_{e,f}^4\,\epsilon_{B,f}^{1/2}\cr
&&\hspace{4.3cm}\times\,A^{-1/2}_{\star,f}\,E_{54.3}\,  t_{1}^{-2}\cr
E^{ssc}_{\rm \gamma, c,f}  &\simeq&  7.7\times 10^{-10}\, {\rm eV}\, \biggl(\frac{1+z}{1.34}\biggr)^{-3}\xi_{-0.3}^{12}\,(1+x_f)^{-4}\,\epsilon_{B,f}^{-7/2}\cr
&&\hspace{4.7cm}\times\, A^{-9/2}_{\star,f}\,E_{54.3}\, t_{1}^{2}\,. \cr
\eary
}
From eqs. (\ref{rho}) and  (\ref{Fsyn}), the maximum SSC flux {\small $F^{ssc}_{\rm \gamma,max,f}\simeq\,(\sigma_T/m_p)\,r\,\rho\,F^{syn}_{\rm \gamma,max,f}$} can be explicity written as
{\small
\bary
F^{ssc}_{\rm \gamma,max,f}&\simeq&5.8 \times 10^5 \,{\rm mJy} \biggl(\frac{1+z}{1.34}\biggr)^2\,\epsilon_{B,f}^{1/2} \,A^{5/2}_{\star,f}\,D^{-2}_{28}\,t^{-1}_1.
\eary
}
In the Klein-Nishina (KN) regime, the emissivity of IC radiation per electron is independent of the electron energy and reduced in comparison with the classical regime, hence the break energy in KN regime  is
{\small
\bary
E^{KN}_{\rm \gamma,f}&\simeq& 1.5\times 10^{-4}\, {\rm GeV}\, (1+x_f)^{-1}\,\biggl(\frac{1+z}{1.34}\biggr)^{-2}\xi_{-0.3}^{4}\epsilon^{-1}_{B,f}\,\Gamma^2_{\star,f}\cr
&&\hspace{5cm}\times\,A^{-1}_{\star,f}\,t_1\,.
\eary
}
From the SSC break energies (eq. \ref{sscforw_a}) and Compton spectra \citep{2015ApJ...804..105F}, the LC in the fast- and slow-cooling regime can be written as
{\small
\begin{eqnarray}
\label{fcic_t}
[F_\nu]^{ssc} \propto\cases{ 
t^0,\hspace{1cm}E^{ssc}_{\rm \gamma,c,f}<E^{ssc}_\gamma<E^{ssc}_{\rm \gamma,m,f}, \cr
t^{-p+1},\hspace{0.4cm}E^{ssc}_{\rm \gamma,m,f}<E^{ssc}_\gamma < E^{ssc}_{\rm \gamma,max,f}\,, \cr
}
\end{eqnarray}
}
and
{\small
\begin{eqnarray}
\label{scic_t}
[F_\nu]^{ssc}\propto\cases{ 
t^{-p},\hspace{0.7cm}E^{ssc}_{\rm \gamma,m,f}<E^{ssc}_\gamma<E^{ssc}_{\rm \gamma,c,f},\cr
t^{-p+1},\hspace{0.4cm} E^{ssc}_{\rm \gamma,c,f}<E^{ssc}_\gamma  < E^{ssc}_{\rm \gamma,max,f}\,,\cr
}
\end{eqnarray}
}
respectively.
\subsection{Reverse Shocks}
For the RS, a simple analytic solution can be derived taking two limiting cases, thick- and thin-shell case,  \citep{1995ApJ...455L.143S} by using a critical Lorentz factor ($\Gamma_c$) which is defined by 
\bary\label{gamma_cr}
\Gamma_c&\simeq&134.1\biggl(\frac{1+z}{1.34}\biggr)^{1/4}\, \xi_{-0.3}^{-1/2} \,A^{-1/4}_{\star,r}\,E^{1/4}_{54.3}T_{90,2}^{-1/4}\,,
\eary
where  $T_{90}$ is the duration of the prompt phase and  $A_{\star,r}=A_r/\,(5.0\times 10^{11}\, \rm{g/cm})$ \citep{2000ApJ...536..195C, 2013ApJ...776..119L}.  The synchrotron  spectral evolution between RS and FS is related by 
\bary\label{conec}
E^{syn}_{\rm \gamma, m,r}(t_d)&\sim&\,\mathcal{R}^2_e\,\mathcal{R}^{-1/2}_B\,\mathcal{R}^{-2}_M\,E^{syn}_{\rm \gamma,m,f}(t_d)\cr
E^{syn}_{\rm \gamma,c,r}(t_d)&\sim&\,\mathcal{R}^{3/2}_B\,\mathcal{R}^{-2}_x\,E^{syn}_{\rm \gamma,c,f}(t_d)\cr
F^{syn}_{\rm \gamma,max,r}(t_d)&\sim&\,\mathcal{R}^{-1/2}_B\,\mathcal{R}_M\,F^{syn}_{\gamma,max,f}(t_d)\,,
\eary
where
{\small
\be\label{param}
\mathcal{R}_B=\frac{\epsilon_{B,f}}{\epsilon_{B,r}},\hspace{0.1cm} \mathcal{R}_e=\frac{\epsilon_{e,r}}{\epsilon_{e,f}},\hspace{0.1cm}  \mathcal{R}_x=\frac{1+x_f}{1+x_r+x^2_r}\hspace{0.1cm} {\rm and}\hspace{0.1cm} \mathcal{R}_M=\frac{\Gamma^2_{d}}{\Gamma_r}\,,
\ee
}
and  $\Gamma_{d}\sim$ min ($\Gamma_r,\, 2\Gamma_c$) is the bulk Lorentz factor at the shock crossing time  $t_d\sim \left(\frac{\Gamma_d}{\Gamma_c}\right)^{-4}\,T_{90}$  and $\Gamma_r= \Gamma_{\star,r}\,10^2$ is the bulk Lorentz factor of RS \citep{2003ApJ...595..950Z,2007ApJ...655..973K}. The previous relations tell us that including the re-scaling there is a unified description between forward and reverse shocks, and the distinction between forward and reverse magnetic fields considers that in some central engine models  \citep{1992Natur.357..472U, 1997ApJ...482L..29M,2000ApJ...537..810W} the fireball could be endowed with '"primordial" magnetic fields.  
The RS becomes relativistic during its propagation and the ejecta is significantly decelerated.  The bulk Lorentz factor at the shock crossing time  $t_d\leq T_{90}$ is given by the condition $\Gamma_{r} > 2 \Gamma_c$.  Eventually,  the shock crossing time could be much shorter than $T_{90}$ depending on the degree of magnetization of the ejecta, defined as the ratio of Poynting flux to matter energy flux  $\sigma =L_{pf}/L_{kn}\sim \epsilon_{B,r}$  \citep{2004A&A...424..477F,2005ApJ...628..315Z,2007ApJ...655..973K}. \\
\paragraph{Synchrotron emission}. Assuming that electrons are accelerated in the RS to a power-law distribution and the energy density is equipartitioned  between electrons and the magnetic field, then the e-minimum Lorentz factor and the magnetic field are 
{\small
\be\label{gamma_mr}
\gamma_{\rm e,m,r}=116.2  \biggl(\frac{1+z}{1.34}\biggr)^{-1/4}\xi_{-0.3}^{1/2}\,\epsilon_{e,r}\,  \Gamma_{\star,r}A^{1/4}_{\star,r}\,E^{-1/4}_{54.3} t^{1/4}_{d,1},
\ee
}
and
\bary\label{Br}
B'_r&\simeq& 6.6 \times 10^3\, {\rm G}  \biggl(\frac{1+z}{1.34}\biggr)^{1/2}\xi_{-0.3}^{-1}\, \epsilon_{B,r}^{1/2}\,\Gamma_{\star,r}\,E^{-1/2}_{54.3}\,t_{d,1}^{-1/2}\cr
&&\hspace{5.6cm}\times\,A_{\star,r}\,,
\eary
respectively.  Comparing  the dynamical, cooling  and the acceleration time scales as showed for FS, we can obtain the cooling and maximum Lorentz factors.  By considering $\gamma_{e,a}\simeq\gamma_{e,m}$ \citep{2001ApJ...548..787S} and from eq. (\ref{param}), we re-scale the synchrotron self-absorption energy between FS and RS as {\small $E^{syn}_{\rm \gamma, a,r}\sim\,\mathcal{R}^2_e\,\mathcal{R}^{-1/5}_B\,\mathcal{R}^{-2}_M\,E^{syn}_{\rm \gamma,a,f}$}. From eqs. (\ref{synforw_a}),  (\ref{conec}) and (\ref{param}), we get the synchrotron spectral breaks  
{\small
\bary\label{synrev_a}
E^{syn}_{\rm \gamma,a,r}&\simeq& 1.3\times 10^{-6} \, {\rm eV}\, \biggl(\frac{1+z}{1.34}\biggr)^{-7/5}\xi_{-0.3}^{4/5}\,\epsilon_{e,r}^{-1}\,\epsilon_{B,r}^{1/5}\,\Gamma^{2}_{\star,r}\cr
&&\hspace{4.0cm}\times A_{\star,r}^{11/5}\,E^{-7/5}_{54.3}\,t_{d,1}^{2/5} \cr
E^{syn}_{\rm \gamma,m,r}&\simeq& 1.1 \times 10^2  \, {\rm eV}\, \biggl(\frac{1+z}{1.34}\biggr)^{-1/2}\xi_{-0.3}^{-1}\,\epsilon_{e,r}^{2}\,\epsilon_{B,r}^{1/2}\,\Gamma^{2}_{\star,r}\cr
&&\hspace{4.0cm}\times A_{\star,r}\,E^{-1/2}_{54.3}\,t_{d,1}^{-1/2} \cr
E^{syn}_{\rm \gamma,c,r}&\simeq& 3.7\times 10^{-8} \, {\rm eV}\,  \biggl(\frac{1+z}{1.34}\biggr)^{-3/2}\xi_{-0.3}^{5}\,(1+x_r+x^2_r)^{-2}\cr
&&\hspace{3.2cm}\times\, \epsilon_{B,r}^{-3/2}\,A^{-2}_{\star,r}\,E^{1/2}_{54.3}\,t_{d,1}^{1/2} \cr
F_{\rm \gamma,max,r}&\simeq&  4.1\times10^7  \,{\rm \,mJy}\,  \biggl(\frac{1+z}{1.34}\biggr)^{2}\xi_{-0.3}^{-2}\,\epsilon_{B,r}^{1/2}\,\Gamma^{-1}_{\star,r}\,A^{1/2}_{\star,r}\cr
&&\hspace{4.0cm}\times \,D^{-2}_{28}\,E_{54.3}\,t_{d,1}^{-1}\,.
\eary
}
Synchrotron LCs are derived in \cite{2000ApJ...545..807K}.  Relativistic electrons accelerated at RS radiate photons in optical wavelengths.  Firstly,  synchrotron flux  increases proportionally to $\sim t^{1/2}$, being able to reach a peak time at  $t_d\sim \left(\frac{\Gamma_d}{\Gamma_c}\right)^{-4}\,T_{90}$. The optical flux at the peak can be written as   
{\small
\bary\label{synpeak}
F^{syn}_{\rm \gamma,peak,r}&\simeq&  1.3\times 10^5  {\rm  mJy}\,  \biggl(\frac{1+z}{1.34}\biggr)^{5/4}\,(1+x_r+x^2_r)^{-1}\,\xi_{-0.3}^{1/2}   \epsilon_{B,r}^{-1/4}\,\cr
&&\hspace{0.6cm}\times\,\Gamma^{-1}_{\star,r}\,A_{\star,r}^{-1/2} D_{28}^{-2}\, E^{5/4}_{54.3}\,t_{d,1}^{-3/4}\,  \left(\frac{E^{syn}_{\rm \gamma,r}}{2\,{\rm eV}}\right)^{-1/2}\,.
\eary
}
After that the synchrotron flux  starts decreasing as $\sim t^{-3}$ \citep{2003ApJ...597..455K}.
\paragraph{SSC emission}. Accelerated electrons can upscatter photons from low to high energies as
{\small
\bary\label{ic}
&&E^{\rm ssc}_{\rm \gamma,a,r}\sim2\gamma^2_{e,m,r}E^{syn}_{\rm \gamma, a,r},\hspace{1cm} E^{\rm ssc}_{\rm \gamma,m,r}\sim2\gamma^2_{e,m,r}E^{syn}_{\rm \gamma, m,r},\cr
&&E^{ssc}_{\rm \gamma,c,r}\sim2\gamma^2_{\rm e,c,r}\,E^{syn}_{\gamma,c,r},\hspace{0.3cm} {\rm and}\hspace{0.3cm}F^{ssc}_{\rm \gamma,max,r}\sim\,k\tau\,F^{syn}_{\rm \gamma,max,r}\,,
\eary
}
where {\small $k=4(p-1)/(p-2)$} and {\small $\tau=\frac{\sigma_T N(\gamma_e)}{4\pi r_d}$} is the optical depth of the shell. From eqs. (\ref{synrev_a}) and (\ref{ic}), we get the break SSC energies     
{\small
\bary\label{ssc_a}
E^{ssc}_{\rm \gamma,m,r}&\simeq& 2.1 \times 10^3 \, {\rm \ MeV}\,  \biggl(\frac{1+z}{1.34}\biggr)^{-1}\,\epsilon_{e,r}^4\,\epsilon_{B,r}^{1/2}\,\Gamma^{4}_{\star,r}\cr
&&\hspace{4.8cm}\times  A^{3/2}_{\star,r}\,E^{-1}_{54.3},\cr
E^{ssc}_{\rm \gamma,c,r}&\simeq&  6.5\times 10^{-8} \, {\rm \ eV}\, \biggl(\frac{1+z}{1.34}\biggr)^{-3/2}\xi_{-0.3}^{9}\,(1+x_r+x^2_r)^{-4}\cr
&&\hspace{2.6cm}\times \epsilon_{B,r}^{-7/2}\,\Gamma^{-6}_{\star,r}\,A^{-6}_{\star,r}\,E^{5/2}_{54.3}\,t_{d,1}^{1/2},\cr
F^{ssc}_{\rm \gamma,max,r}&\simeq& 9.3\times10^7 \,{\rm \,mJy}\, \biggl(\frac{1+z}{1.34}\biggr)^{3}\xi_{-0.3}^{-4}\,\epsilon_{B,r}^{1/2}\,\Gamma^{-2}_{\star,r}\,A^{3/2}_{\star,r}\cr
&&\hspace{3.9cm}\times D^{-2}_{28}\,E_{54.3}\,t_{d,1}^{-2}\,,
\eary
}%
and the break energy at the KN regime is
{\small
\be\label{kn}
E^{KN}_{\gamma,r}\simeq 4.3\times 10^{-3}\,  {\rm GeV}(1+x_r+x^2_r)^{-1}\xi_{-0.3}^{2}\,\epsilon_{B,r}^{-1}\,E_{54.3}\,\Gamma^{-2}_{\star,r}\,A^{-2}_{\star,r}\,.
\ee
}
LC of Compton scattering emission can be analytically derived from  \citet{2000ApJ...536..195C}.  For  $t < t_d$,  we take into account that: i) the maximum synchrotron flux  is a constant function of time $F^{syn}_{\rm \gamma, max}\sim t^0$, ii) the spherical radius and the number of radiating electrons  in the shocked shell region increase with time as $r\sim t$ and N$(\gamma_e)\sim t$, respectively,  and iii) the cooling break energy  $E^{ssc}_{\rm \gamma, c}\sim \gamma^2_c  E^{syn}_{\rm \gamma, c}$ increases as $\sim t^3$, then the SSC flux increases as $F^{ssc}_{\rm \nu}\sim {E^{ssc}_{\rm \gamma, c}}^{1/2} F^{ssc}_{\rm \nu, max} \sim {E^{ssc}_{\rm \gamma, c}}^{1/2} \frac{N_e}{r_d^2} F^{syn}_{\rm \gamma, max}\sim t^{1/2}$.   For $t  > t_d$, we consider that the characteristic break energy $E^{ssc}_{\rm \gamma, m}\sim \gamma^2_m  E^{syn}_{\rm \gamma, m}$ decreases as $\sim t^{-1}$, then the SSC flux decreases as $F^{ssc}_{\rm \nu}\sim {E^{ssc}_{\rm \gamma, m}}^{-(p-1)/2} F^{ssc}_{\rm \nu, max} \sim {E^{ssc}_{\rm \gamma, m}}^{-\frac{p-1}{2}} \frac{N_e}{r_d^2} F^{syn}_{\rm \gamma, max}\sim t^{-\frac{p+1}{2}}$.  It is worth noting that the decay index of the emission for  $t  > t_d$ might be higher than $\frac{p-1}{2}$ due to the angular time delay effect \citep{2003ApJ...597..455K}.
At $t_d$, the $\gamma$-ray  flux peaks at \citep{2003ApJ...597..455K}:
{\small
\bary\label{sscpeak}
F^{ssc}_{\rm \gamma,peak,r}&\simeq& 6.7\times 10^{-6}  {\rm  mJy}  \left(\frac{1+z}{1.34}\right)^{-1/2}\xi_{-0.3}^9\,x_r\,(1+x_r+x^2_r)^{-5}\cr
&&\hspace{1.2cm}\times\,\epsilon_{e,r}\,\epsilon_{B,r}^{-7/2}\,\Gamma^{-6}_{\star,r} \,A^{-6}_{\star,r}\, D^{-2}_{28}\,E_{54.3}\,t_{d,1}^{-1/2}\cr
&&\hspace{3.5cm}\times \left(\frac{E^{ssc}_{\rm \gamma,r}}{100\,{\rm MeV}}\right)^{-1/2}.
\eary
}
\section{Discussion}
\begin{figure}
\epsscale{0.8}
\plotone{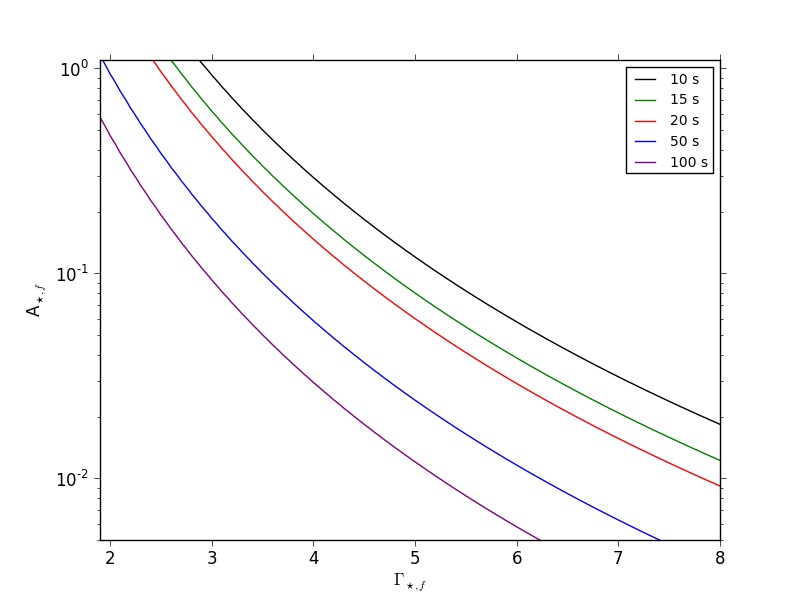}
\caption{Contour lines of A$_{\star,f}$ and bulk Lorentz factor ($\Gamma_{\star,f}$) as a function of the deceleration time $t_{dec}$.  We use four values of deceleration times $t_{dec}=$ 10, 15, 20, 50 and 100 s}
\label{con_lin}
\end{figure}

\begin{figure}
\epsscale{1.2}
\plotone{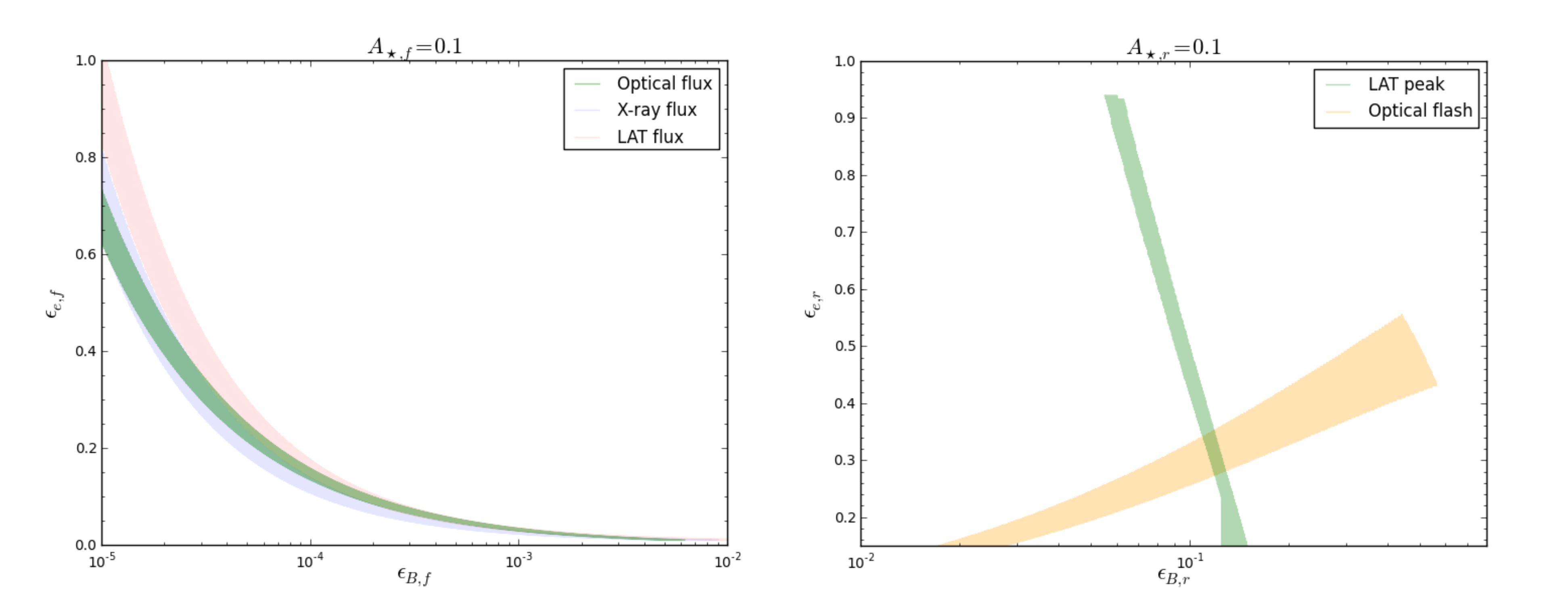}
\caption{Values of equipartition parameters  for the FS (right) and RS (left) that reproduce the multiwavelength afterglow observed in GRB 130427A. }
\label{parameter}
\end{figure}

\begin{figure}
\epsscale{0.8}
\plotone{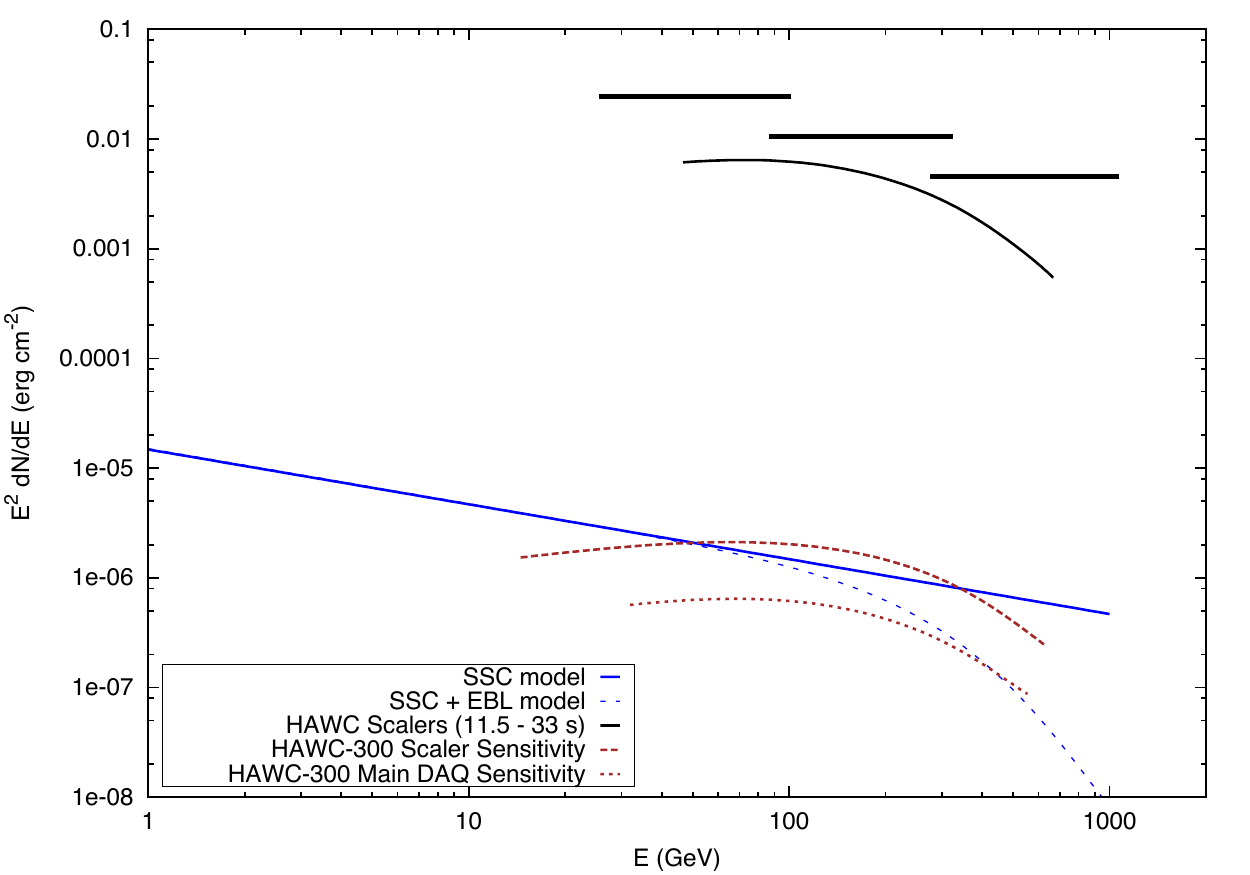}
\caption{The SSC model proposed in this work and upper limits placed by the HAWC Observatory in the time interval [11.5 - 33 s].  Blue lines show the SSC model without (continuous) and with the effect of the EBL absorption (dashed).  Black solid lines display the scaler limit. Brown dashed and dotted lines exhibit the sensitivity of the two HAWC DAQs for the full detector. (For details see  \cite{2015ApJ...800...78A}).} 
\label{hawc_limit}
\end{figure}

Recently, \citet{2015ApJ...804..105F} presented a leptonic model based on an early afterglow that evolved in a stellar wind to describe successfully the  multiwavelength afterglow observations of GRB110731A.  In this work,  we have used this model to explain the  multiwavelength afterglow observed in GRB 130427A.  Requiring the values of isotropic energy $E\simeq 1.4\times 10^{54}$ erg \citep{2014Sci...343...38V} and the redshift  $z=0.34\pm0.01$\citep{2013GCN..14686...1L, 2013GCN..14491...1F}, from eq. (\ref{t_dec}) we have plotted the contour lines of the external medium density and bulk Lorentz factor for five values of the deceleration time $t_{dec}=$10, 15, 20, 50 and 100 s, as shown in Figure \ref{con_lin}. Taking into account the fact that the LAT peak and the bright optical flash were present and also showed a close correlation in the time interval [9.31 s,19.31 s] \citep{2014Sci...343...38V},  we have considered the values of the external medium density and bulk Lorentz factor for which the deceleration time (eq. \ref{t_dec}) is t$_{dec}$=10 s  (line in black color). \\
To obtain the values  of densities ($A_{\star,f/r}$) and  the equipartition parameters ($\epsilon_{B,f/r}$ and $\epsilon_{e,f/r}$)  that reproduce the multiwavelength afterglow observed in GRB 130427A,  we have used the method of Chi-square $ \chi^2$ minimization as implemented in the ROOT software package \citep{1997NIMPA.389...81B}.  The LAT flux has been fitted  by synchrotron radiation from FS and SSC emission from RS; the whole temporally extended emission  using synchrotron LC in the  fast-cooling regime (eq. \ref{A1})  and the bright peak  at 15 s  with  SSC emission (eq. \ref{sscpeak}),  for high-energy electrons emitting photons at 100 MeV. The X-ray flux has been fitted with the synchrotron LC in the slow-cooling regime for electrons radiating at  $E^{syn}_{\gamma,f}=$ 10 keV (eq. \ref{A2}), and the optical flux has been described by synchrotron radiation from FS and RS;  the  temporally extended emission using the LC of FS in the  slow-cooling regime (eq. \ref{A3}) and the bright optical flash  with the  LC of RS in fast-cooling regime (eq. \ref{synpeak}) for $E^{syn}_{\gamma,r} =$ 2 eV.\\ 
Figure \ref{parameter} (left panel) visualizes the values of equipartition parameters, $\epsilon_{B,f}$ and  $\epsilon_{e,f}$ for $A_{\star,f}=10^{-1}$, that reproduce the temporally extended emissions of LAT, X-ray and optical data (see Figure  \ref{fit_afterglow}).   It displays  areas in red, blue and green colors.  The area in red  exhibits the set of parameters that describes the extended LAT component, the area in blue  displays those parameters that describe the X-ray emission and the area in green the parameters that describe the extended optical flux.  Regions where the areas intercept correspond to the set of parameters that reproduces more than one flux at the same time.  As shown, the set of parameter values: $\epsilon_{e,f}\sim$ 0.32 and $\epsilon_{B,f}\,\sim 3\times 10^{-5}$ for $A_{\star,f}=10^{-1}$ and $\Gamma_{\star,f}\simeq5.5$,    can reproduce the temporally extended LAT, X-ray and optical fluxes. \\
Figure \ref{parameter} (right panel) visualizes the values of equipartition parameters, $\epsilon_{B,r}$ and  $\epsilon_{e,r}$ for $A_{\star,r}= 10^{-1}$, that describe the bright LAT peak and the optical flash (see Figure  \ref{fit_afterglow}). It shows areas in green and yellow. The area in green  exhibits the set of parameters that describes the LAT peak and the area in yellow displays those parameters that explain the bright optical flash. It can be seen that the set of parameter values:  $\epsilon_{e,r}\sim$ 0.32 and $\epsilon_{B,r}\,\sim 0.13$ for $A_{\star,r}=10^{-1}$ and $\Gamma_{\star,r}\simeq 5.5$,  is able to generate the bright LAT peak and the optical flash.\\
Figure \ref{fit_afterglow} shows the contributions of synchrotron radiation from FS (dashed lines) and RS (dotted-dashed line), and SSC (continuous line) emission from RS  to the multiwavelength afterglow observed in GRB 130427A.\\
In Table 2, we summarize the equipartition parameters, densities and bulk Lorentz factors found after fitting the multiwavelength afterglow observed in GRB130427A. In addition,  the parameters obtained for GRB 110731A have been included in order to compare them with those obtained in GRB 130427A.
\begin{center}
\begin{center}
\scriptsize{\textbf{Table 2. Parameters found after fitting the multiwavelength afterglow observations of GRB110731A and GRB130427A.}}
\end{center}
\begin{tabular}{ l c c }
  \hline \hline
 \scriptsize{} & \scriptsize{GRB110731A} & \scriptsize{GRB130427A} \\
 \hline
 Forward shock\\
\hline
\scriptsize{$\epsilon_{B,f}$}    & \scriptsize{ $7\times 10^{-5}$ }  &  \scriptsize {$3\times 10^{-5}$} \\
\scriptsize{$\epsilon_{e,f}$}    & \scriptsize{0.4}                         &  \scriptsize{0.32} \\
\scriptsize{$A_{f}\, (5\times 10^{11}\,{\rm g/cm})$ }    & \scriptsize{$10^{-1}$}  &  \scriptsize {$10^{-1}$} \\
\scriptsize{$\Gamma_f$}    & \scriptsize{ 520}  &  \scriptsize {550} \\
 \hline
 Reverse shock\\
 \hline
\scriptsize{$\epsilon_{B,r}$}    & \scriptsize{ 0.28}  &  \scriptsize {$0.13$} \\
\scriptsize{$\epsilon_{e,r}$}    & \scriptsize{0.4}  &  \scriptsize {0.32} \\
\scriptsize{$A_{r}\, (5\times 10^{11}\,{\rm g/cm})$}    & \scriptsize{$10^{-1}$}  &  \scriptsize {$10^{-1}$} \\
\scriptsize{$\Gamma_r$}    & \scriptsize{ 520}  &  \scriptsize {550} \\
 \hline
\end{tabular}
\end{center}
\begin{center}
\end{center}
The set of parameter values  obtained using our model is similar to those used to successfully describe  the afterglow observed at different times \citep{2013ApJ...776..119L, 2014ApJ...781...37P, 2014Sci...343...42A}.   Comparing the equipartition parameters in both shocks,  it is possible to observe that the energy fraction going into electron acceleration is equal ($\mathcal{R}_e=1$) and the magnetic fields in both shocks (eqs. \ref{Bf} and \ref{Br}) are different  $B'_r=\mathcal{R}^{-1/2}_B \,B'_f= 65.8\,B'_f$.\\
Putting together the parameters for both GRB130427A and GRB110731A, a strong similarity  can be noted between them.  For instance, the value of the bulk Lorentz factor found for GRB130427A lies not only in the range of GRB 110731A but also in the similar range of values ($\Gamma\sim$ 500 - 600) demanded for most LAT-detected and high-redshift  GRBs \citep{2012ApJ...755...12V}.\\ 
Using the values of parameters reported in Table 2 and eqs. (\ref{scsyn_t}), (\ref{sscforw_a}), (\ref{synrev_a}), (\ref{ssc_a}), the observable quantities have been computed, as shown in Table 3. Again, we have put together the observable quantities obtained for GRB 110731A.\\
The maximum photon energy achieved by synchrotron radiation is $E^{syn}_{\rm \gamma,max,f}$= 107.7 (60.6) GeV for t=$10\, (10^2)$ s. Therefore, the highest-energy photon of $95$ GeV at 244 s after the GBM trigger cannot be generated from synchrotron radiation in the standard afterglow model.   The highest-energy photon could be interpreted in the SSC framework, for which the VHE flux is expected to peak at $\simeq$ 22 TeV (see Table 3).  However,  high-energy photons were searched for in this burst by the HAWC observatory, and although no significant excess of counts was observed,  upper limits were placed \citep{2015ApJ...800...78A}.  Figure \ref{hawc_limit} shows the integral flux upper limits placed by HAWC observatory in the time interval [11.5 - 33 s] and the SSC emission without (continuous line) and with (dashed line) the effect of the EBL absorption (dashed;  \cite{2008A&A...487..837F}). As shown, the SSC flux (less than 1 TeV)  is low enough to be observed by HAWC when it was running at 10\% of the final detector.  Therefore, with the parameters found after fitting the multiwavelength afterglow (see table 2)  not only the highest-energy photon  but also the VHE photon non-detection of GRB 130427A could be interpreted in the SSC framework.\\
\begin{center}
\begin{center}
\scriptsize{\textbf{Table 3. Quantities obtained  with our model for  GRB110731A and GRB130427A}}\\
\end{center}
\begin{tabular}{ l c c }
  \hline \hline
 \scriptsize{} & \small{GRB110731A} & \small{GRB130427A} \\
 \hline\hline
 Forward shock\\
\hline\hline
\scriptsize{$t_{dec}$ (s)} & \scriptsize{$5.6$} & \scriptsize{$9.9$}  \\
\scriptsize{$B'_f$ (G)} &\scriptsize{$52.7$} &\scriptsize{$18.9$}\\
\hline
\small{Synchrotron emission}\\
\hline
\scriptsize{$E^{syn}_{\rm \gamma,a,f}$ (eV)}    & \scriptsize{$5.6\times 10^{-4}$}  &  \scriptsize{$3.3\times 10^{-2}$} \\
\scriptsize{$E^{syn}_{\rm \gamma,m,f}$ (keV)}    & \scriptsize{$77.5$}  &  \scriptsize{$23.1$} \\
\scriptsize{$E^{syn}_{\rm \gamma,c,f} $} (eV)   & \scriptsize{0.30}  &  \scriptsize{1.3} \\
\scriptsize{$E^{syn}_{\rm \gamma,max,f}$ (GeV)}    & \scriptsize{$36.9$}  &  \scriptsize{$107.7$} \\
\hline
\small{SSC emission}\\
\hline
\scriptsize{$E^{ssc}_{\rm \gamma,m,f}$ (TeV)}    & \scriptsize{$11.7$}  &  \scriptsize{$22.1$} \\
\scriptsize{$E^{ssc}_{\rm \gamma,c,f}$ (TeV)}    & \scriptsize{$8.4\times 10^{-8}$}  &  \scriptsize{$1.4\times 10^{-7}$} \\
\scriptsize{$E^{KN}_{\rm \gamma,f}$ (TeV)}    & \scriptsize{$42.3\times 10^{-3}$}  &  \scriptsize{$102.3\times 10^{-3}$} \\
 \\
 \hline\hline
 Reverse shock\\
\hline\hline
\scriptsize{$\Gamma_c$} & \scriptsize{472.5} &\scriptsize{236.7} \\
\scriptsize{$B'_r$ (G)} &  \scriptsize{$3.8\times 10^3$}& \scriptsize{$1.7\times 10^3$}\\
\hline
\small{Synchrotron emission}\\
\hline
\scriptsize{$E^{syn}_{\rm \gamma,a,r}$ (eV)}    & \scriptsize{$4.3\times 10^{-8}$}  &  \scriptsize{$0.5\times 10^{-7}$} \\
\scriptsize{$E^{syn}_{\rm \gamma,m,r}$ (eV)}    & \scriptsize{$128.9$}  &  \scriptsize{$14.3$} \\
\scriptsize{$E^{syn}_{\rm \gamma,c,r}$ (eV)}    & \scriptsize{$0.9\times 10^{-5}$ }  &  \scriptsize{$2.5\times 10^{-5}$} \\
\hline
\small{SSC emission}\\
\hline
\scriptsize{$E^{ssc}_{\rm \gamma,m,r}$ (MeV)}    & \scriptsize{$1.1\times 10^2$}  &  \scriptsize{$0.6\times 10^2$} \\
\scriptsize{$E^{ssc}_{\rm \gamma,c,r}$ (eV)}    & \scriptsize{$5.9\times 10^{-3}$}  &  \scriptsize{$1.4\times 10^{-5}$} \\
\scriptsize{$E^{KN}_{\rm \gamma,r}$ (GeV)}    & \scriptsize{$52.7 $}  &  \scriptsize{$166.2$} \\
\hline
\end{tabular}
\end{center}
%
%
%
%
The values of deceleration time and critical Lorentz factor computed in our model are self-consistent with the fact that both the bright LAT peak and the optical flash take place in the time interval [9.31 s,19.31 s] peaking at 15 s and the RS evolves in the thick-shell case ($\Gamma_r> 2\Gamma_c$).\\
 The synchrotron self-absorption energies  from FS and RS are in the weak self-absorption regime, then, as observed in LC of GRB 130427A, there is no thermal peak in the synchrotron spectrum due to pile-up electrons \citep{2004ApJ...601L..13K,2013MNRAS.435.2520G}. \\
Unlike GeV, X-ray and optical early observations, GRB130427A started to be observed at $\sim$ 0.3 days  in radio wavelengths. This burst was followed  for more than 4 months by Westerbork Synthesis Radio Telescope (WSRT), European Very Long Baseline Interferometry Network (EVN), Combined Array for Research in Millimeter Astronomy (CARMA), Very Large Array (VLA) and other radio observatories \citep{2014MNRAS.444.3151V, 2014ApJ...781...37P}.  In particular, the observable quantities of radio observations  at 15 GHz are given in Table 4 \citep{2014MNRAS.444.3151V}.
 \begin{center}
\begin{center}
\scriptsize{\textbf{Table 4. Temporal power-law indices and fluxes of radio observation at 15 GHz \citep{2014MNRAS.444.3151V}.}}\\
\end{center}
\begin{tabular}{ l c c }
  \hline \hline
 \small{Temporal index} & \small{Time range} & \small{Flux} \\
 \small{} & \small{(days)} & \small{(mJy)}\\
\hline\hline
 \scriptsize{$0.33\pm 0.20$} & \scriptsize{0.3 - 0.7} & \scriptsize{$\sim$ 3.6}  \\
\scriptsize{$-1.16 \pm 0.14  $} &\scriptsize{0.7 - 4}  & \scriptsize{$\sim$ 1.1}\\
\scriptsize{$-0.48 \pm 0.07 $} &\scriptsize{4 - 60}    & \scriptsize{$\sim$ 0.2}\\
\hline
\hline
\end{tabular}
\end{center}
Following \cite{2013NewAR..57..141G}, we derive the LC of synchrotron radiation from FS in the radio frequencies.  The synchrotron spectrum in the radio frequencies is
{\small
\begin{eqnarray}
\label{fcsyn_radio}
[F_\nu]^{syn}&=&F^{syn}_{\rm \gamma,max,f}\cr
&&\times\cases{
\left(\frac{E^{syn}_{\rm \gamma,f}}{E^{syn}_{\rm \gamma,m,f}}\right)^{1/3}{\rm for}\,\, E^{syn}_{\rm \gamma,a,f}< E^{syn}_{\rm \gamma,f}< E^{syn}_{\rm \gamma,m,f}<E^{syn}_{\rm \gamma,c,f}, \cr
\left(\frac{E^{syn}_{\rm \gamma,f}}{E^{syn}_{\rm \gamma,m,f}}\right)^{-\frac{p-1}{2}}{\rm for}\,\, E^{syn}_{\rm \gamma,a,f}< E^{syn}_{\rm \gamma,m,f}< E^{syn}_{\rm \gamma,f}<E^{syn}_{\rm \gamma,c,f}. \cr
}
\end{eqnarray}
}
From eqs. (\ref{t_dec}), (\ref{synforw_a}) and (\ref{Fsyn}), we get that the maximum synchrotron flux and the characteristic break synchrotron energy as a function of time are  {\small $F^{syn}_{\rm \gamma,max,f}\propto  \epsilon_{B,f}^{1/2} \,A^{3/2}_{\star,f}\,D^{-2}\,\Gamma^2_{\star,f}$} and {\small $E^{syn}_{\rm \gamma,m,f}\propto  \epsilon_{e,f}^2\,\epsilon_{B,f}^{1/2} \,A^{1/2}_{\star,f}\,\Gamma^2_{\star,f}\,t^{-1}$}, respectively.  Taking into account the values reported in Table 2 and the synchrotron spectrum \citep{2013NewAR..57..141G}, the LC of synchrotron radiation in the radio frequencies is in the form
{\small
\begin{eqnarray}
\label{fcsyn_radio}
[F_\nu]^{syn}= \cases{
F^{syn}_{\nu,rh},\hspace{0.5cm}\, E^{syn}_{\rm \gamma,a,f}< E^{syn}_{\rm \gamma,f}< E^{syn}_{\rm \gamma,m,f}<E^{syn}_{\rm \gamma,c,f} ,\cr
F^{syn}_{\nu,rs},\hspace{0.5cm}\,E^{syn}_{\rm \gamma,a,f}< E^{syn}_{\rm \gamma,m,f}< E^{syn}_{\rm \gamma,f}<E^{syn}_{\rm \gamma,c,f}, \cr
}
\end{eqnarray}
}
with $F^{syn}_{\nu,rh}$ and $F^{syn}_{\nu,rs}$ given by
{\small
\bary
F^{syn}_{\nu,rh}&\sim&2.6\,{\rm  mJy}\,\left(\frac{1+z}{1.34}\right)\,\xi^{2/3}\,\epsilon_{e,f,0.5}^{-2/3}\, \epsilon_{B,f,-4.5}^{1/3} \,A^{4/3}_{\star,f,-1}\,D_{28}^{-2}\,\cr
&&\hspace{2.2cm} \times\,\Gamma^{4/3}_{\star,f,0.2}\, t_{4.6}^{1/3}\,\left(\frac{E^{syn}_{\rm \gamma,f}}{15\, {\rm GHz}}\right)^{1/3}\,,
\eary
}
and
{\small
\bary
F^{syn}_{\nu,rs}&\sim&0.4 \,{\rm  mJy}\,\left(\frac{1+z}{1.34}\right)\,\xi^{p-1}\,\epsilon_{e,f,0.5}^{p-1}\, \epsilon_{B,f,-4.5}^{\frac{p+3}{4}} \,A^{\frac{p+5}{4}}_{\star,f,-1}\,D_{28}^{-2}\,\cr
&& \hspace{1.8cm}\times\,\Gamma^{p+1}_{\star,f,-0.3}\,t_6^{-\frac{p-1}{2}}\,\left(\frac{E^{syn}_{\rm \gamma,f}}{15\, {\rm GHz}}\right)^{-\frac{p-1}{2}}\,,
\eary
}
respectively.   At  $t=4\times 10^4$ s, the synchrotron self-absorption and  characteristic break energies are  {\small $E^{syn}_{\rm \gamma,a,f}\sim  3.5$} GHz and $E^{syn}_{\rm \gamma,m,f}\sim 142.4$ GHz, respectively, and at $t=10^6$ s, these break energies are $E^{syn}_{\rm \gamma,a,f}\sim 0.5$ GHz and $E^{syn}_{\rm \gamma,m,f}\sim 1.1$ GHz, respectively, then the transition of flux densities between $F^{syn}_{\nu,rh}$ and $F^{syn}_{\nu,rs}$ occurs at t$_m\sim 10^5$ s. Here, $t_m$ is the critical time when the characteristic break energy  crosses the observed energy $E^{syn}_{\rm \gamma,f}\sim {\rm 15 \,GHz}$. Comparing the values given in Table 4 with the temporal power-law indeces and fluxes obtained in eq. (\ref{fcsyn_radio}), for p=2.2 the radio observations provide a consistency check for our early time modeling.\\
As previously calculated, the magnetic field ratio between the reverse and forward shocks is $\frac{B'_r}{B'_f}\simeq 66$. This result strongly suggests that for GRB 130427A  the ejecta must be magnetized, thus altering the temporal and spectral properties in the photon and neutrino spectra  \citep{2013PhRvL.110l1101Z,2014ApJ...787..140F, 2005MNRAS.364L..42F,2004MNRAS.351L..78F,2007MNRAS.378.1043J,2005ApJ...633.1027S}.
\section{Conclusions}
\begin{figure}
\epsscale{0.8}
\plotone{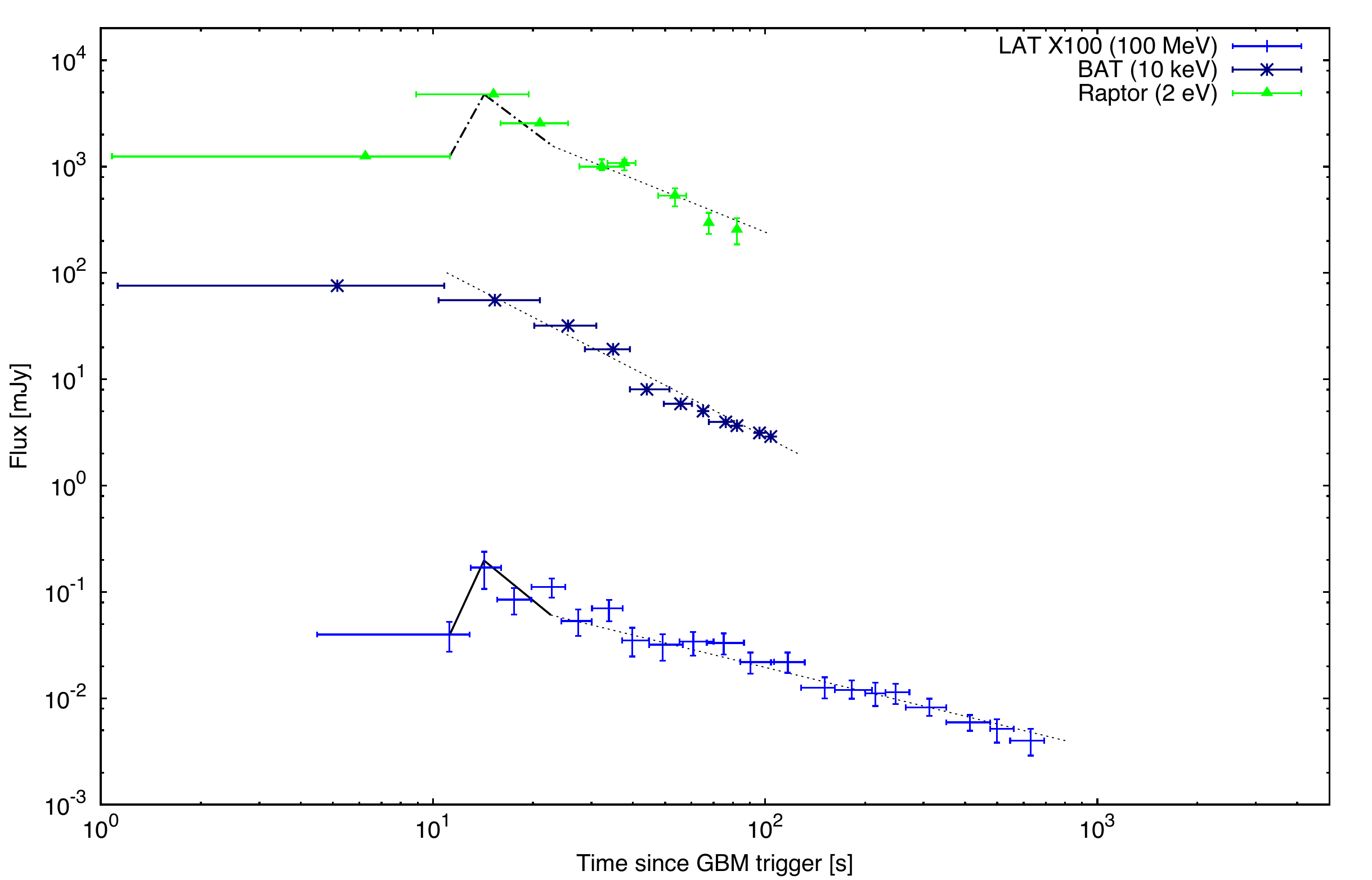}
\caption{Fits of the multiwavelength LCs of GRB 130427A observation with our model. We use the RS  in the thick-shell regime to describe the GeV peak  (continuous line) and optical flash (dash-dotted line) and the FS to explain  the temporally extended LAT, X-ray and optical emissions (dashed lines).}
\label{fit_afterglow}
\end{figure}

%
%
We have  applied the leptonic model previously introduced in \cite{2015ApJ...804..105F} in order to describe the early afterglow emission  of GRB 110731A.  We have modeled the extended LAT, X-ray and optical  emission  by synchrotron emission from FS, and the bright LAT peak and optical flash  by SSC and synchrotron emission from RS, respectively.\\
%
%
We have considered that the ejecta propagating into the stellar wind is decelerated early, at $\sim 10$ s and the RS evolves in the thick-shell regime.   Taking into account  the values for redshift  $z=0.34$, isotropic energy $E\simeq1.4\times 10^{54}$ erg and the stellar wind $A_f=5.0\times 10^{10}$ g/cm, the value of the bulk Lorentz factor as required for most LAT-detected long-duration gamma-ray bursts  lies in the range ($\Gamma\sim$ 500 - 600) \citep{2012ApJ...755...12V, 2013ApJ...763...71A, 2014Sci...343...42A}.\\
To find the values of equipartition parameters $\epsilon_{B,f/r}$ and $\epsilon_{e,f/r}$, we have assumed that the magnetic field and electron parameters are constant and then fittted the multiwavelength afterglow LCs;  the extended temporally emissions (LAT, X-ray and optical) by synchrotron radiation from the FS and the bright LAT peak and optical flash by SSC and synchrotron emission from RS, respectively (see fig. \ref{fit_afterglow}). The values of the parameters found using our model  (see table 2) correspond to those typically used to explain the afterglow observed at different times and energy bands \citep{2013ApJ...776..119L, 2014ApJ...781...37P, 2014Sci...343...42A}. \\ 
The set of parameter values  obtained using our model is similar to those used to describe successfully the afterglow observed at different times.
%
%
Although some authors have claimed that the GeV emission detected by LAT in coincidence with the prompt phase could have an internal origin \citep{2011MNRAS.415...77M, 2011ApJ...730..141Z, 2011ApJ...733...22H,  2011ApJ...730....1L}, this is the first time that the bright LAT peak and the optical flash are observed temporally correlated, being the former event successfully interpreted as RS emission in the early afterglow framework \citep{2014Sci...343...38V}.   Therefore,  it is  overwhelming evidence that the bright LAT peak around the afterglow onset time comes from the RS as has been explained in this work.\\
%
%
%
We have restricted our modeling to the first $\sim 10^3$ s of GRB 130427A when only optical and higher energy observations are available.  For this GRB, radio observations started at $\sim 3\times 10^4$ s. While detailed modeling of the late time emission is outside of the scope of this paper, following \cite{2013NewAR..57..141G} we derive the LC of synchrotron emission from FS in the radio frequencies and  extrapolate our model to $\sim$  GHz at 0.3 $\leq t \leq$ 60 days \citep{2014MNRAS.444.3151V}. Comparing the temporal indeces and fluxes of the radio observations reported in Table 4 with the values obtained in our model (temporal power-law indeces $\alpha=$ 0.33 and -0.6, fluxes $\sim$ 2.6 mJy and 0.4 mJy for $t\lesssim 10^{5}$ s and  $t\gtrsim10^{5}$ s, respectively), we show that our model is consistent to explain the radio observations of this burst.  It is worth noting that the reverse shock contribution in radio is not significant, that is why we calculate the FS flux.\\
%
%
Since  GRB 130427A is the most  powerful burst detected with a $z\leq\,0.5$, copious target photons are expected for photo-hadronic interactions, making them promising candidates for neutrino detection. Searches for high-energy neutrinos in spatial and temporal coincidence around this burst were performed, although no neutrinos were observed \citep{2013GCN..14520...1B}.  As found in this work, the magnetic field in the reverse-shock region is stronger ($\simeq$ 66 times) than in the forward-shock region, indicating that the ejecta of GRB 130427A is magnetized.  The null neutrino result reported by IceCube Collaboration could be explained in the  framework of magnetized outflow where neutrino flux is degraded as was previously pointed out by \cite{2013PhRvL.110l1101Z} and  \cite{2014ApJ...787..140F}.\\ 
%
%
It is worth noting that although any significant excess of counts coming from GRB 130427A has not been observed by the HAWC observatory,  nowadays bursts with identical features can be detected by this TeV $\gamma$-ray observatory. Hence, similar bursts could bring to light information on external medium density, bulk Lorentz factors and energy fractions converted to accelerate electron and/or amplify magnetic fields, thus potentially further constraining possible models.\\
%
%
\acknowledgments
We thank the anonymous referee for a critical reading of the paper and valuable suggestions that helped improve the quality and clarity of this work. We also thank Bing Zhang, Anatoly Spitkovsky, Dimitrios Giannios, Ignacio Taboada and Dirk Lenard for useful discussions. This work was supported by PAPIIT-UNAM IG100414 and Fermi grant NNM11AA01A (PV).
%
%
%

%
\clearpage
\end{document}